\documentclass{article}
\usepackage{amsfonts,amssymb,amsmath,graphicx}
\usepackage{theorem}
\usepackage{caption}
\usepackage{cite}
\usepackage{url}
\usepackage{enumerate}
\usepackage{tikz}
\usepackage[margin=1.2in]{geometry}
\usepackage{todonotes}
\usepackage{color,ulem}

\newcommand{\lda}{\lambda}
\newcommand{\1}{\mbox{\hspace{.0em}1\hspace{-.24em}I}}

\usetikzlibrary{arrows}
\tikzstyle{block}=[draw opacity=0.7,line width=1.4cm]
\usetikzlibrary{positioning,arrows,chains,matrix,scopes,fit,decorations.markings,decorations,decorations.pathreplacing}
\tikzset{
  on each segment/.style={
    decorate,
    decoration={
      show path construction,
      moveto code={},
      lineto code={
        \path [#1]
        (\tikzinputsegmentfirst) -- (\tikzinputsegmentlast);
      },
      curveto code={
        \path [#1] (\tikzinputsegmentfirst)
        .. controls
        (\tikzinputsegmentsupporta) and (\tikzinputsegmentsupportb)
        ..
        (\tikzinputsegmentlast);
      },
      closepath code={
        \path [#1]
        (\tikzinputsegmentfirst) -- (\tikzinputsegmentlast);
      },
    },
  },
  mid arrow/.style={postaction={decorate,decoration={
        markings,
        mark=at position .4 with {\arrow[#1]{stealth}}
      }}},
}

\newtheorem{definition}{Definition}[section]

\newtheorem{proposition}[definition]{Proposition}

\newtheorem{rmk}[definition]{Remark}
\numberwithin{equation}{section}

\newcommand{\mb}[1]{\quad\mbox{#1}\quad}
\newcommand{\beq}{\begin{equation}}
\newcommand{\eeq}{\end{equation}}
\newcommand{\bea}{\begin{eqnarray}}
\newcommand{\eea}{\end{eqnarray}}
\newcommand{\beano}{\begin{eqnarray*}}
\newcommand{\eeano}{\end{eqnarray*}}
\newcommand{\bma}{\begin{pmatrix}}
\newcommand{\ema}{\end{pmatrix}}

\newcommand{\vph}{\varphi}

      \def\cB{{\cal B}}      
            
            \def\cI{{\cal I}}
\def\cJ{{\cal J}}      \def\cK{{\cal K}}      \def\cL{{\cal L}}
\def\cM{{\cal M}}      \def\cN{{\cal N}}

\def\fc{{\mathfrak c}}

\def\ff{{\mathfrak f}}

\newcommand{\CC}{{\mathbb C}}

\newcommand{\KK}{\mbox{${\mathbb K}$}}

\newcommand{\RR}{\mbox{${\mathbb R}$}}

\newcommand{\prf}{\underline{Proof:}\ }
\newcommand{\finprf}{\null \hfill {\rule{5pt}{5pt}}\indent}
\newcommand{\ie}{{\it i.e.}\ }
\newcommand{\cf}{{\it c.f.}\ }
\newcommand{\eg}{{\it e.g.}\ }

\title{Nonlinear Schr\"odinger equation on the half-line\\ without a conserved number of solitons}
\date{\empty}
\author{Vincent Caudrelier$^{1}$, Nicolas Cramp\'e$^{2}$, Eric Ragoucy$^{3}$, 
Cheng Zhang$^{4}$\\
\footnotesize{ v.caudrelier@leeds.ac.uk,
crampe1977@gmail.com, ragoucy@lapth.cnrs.fr, ch.zhang.maths@gmail.com}\\ \\
$^{1}$ \sc \small School of Mathematics, Leeds University, Leeds, LS2 9JT, UK\\
$^{2}$ \sc \small Institut Denis-Poisson CNRS/UMR 7013 - Universit\'e de Tours - Universit\'e
d'Orl\'eans,\\ \sc \small Parc de Grandmont, 37200 Tours, France\\
$^{3}$ \sc \small LAPT{\rm{h}}, CNRS and Universit{\'e} Savoie Mont Blanc, Annecy, 74600, France\\
$^{4}$ \sc \small Department of Mathematics, Shanghai University, Shanghai, 200444, China}

\begin{document}
\maketitle
\begin{abstract}
We explore the phenomena of absorption/emission of solitons by an integrable boundary for the focusing nonlinear Schr\"odinger equation on the half-line.
This is based on the investigation of time-dependent reflection matrices which satisfy the boundary zero curvature equation. In particular, this leads to absorption/emission processes at the boundary that can take place for solitons and higher-order solitons. As a consequence, the usual charges on the half-line are no longer conserved but we show explicitly how to restore an infinite set of conserved quantities by taking the boundary into account.
The Hamiltonian description and Poisson  structure of the model are presented, which allows us to derive for the first time a classical version of the boundary algebra used originally in the context of the quantum nonlinear Schr\"odinger equation.
   \vspace{.2cm}


\vspace{.2cm}

\end{abstract}

\section{Introduction}

The topic of integrable boundary conditions with the objective of constructing initial-boundary value problems that can be solved by taking advantage of the integrability of the Partial Differential Equation (PDE) under study has a long history. The first instance can be found in \cite{AS} where the Inverse Scattering Method (ISM) \cite{GGKM,ZS,AKNS} was combined with the idea of odd-even extensions familiar in linear PDEs to derive solutions of certain integrable PDEs on the half-line with Dirichlet or Neumann boundary conditions. Two important steps were taken later by Sklyanin and Habibullin to propose {\it systematic} methods for: 1) Identifying boundary conditions that preserve (Liouville) integrability \cite{SKBC,sklyanin1988boundary}; 2) Constructing solutions in the spirit of the odd-even extensions by mapping the half-line problem to a full-line \cite{HH1} (see also \cite{BT1, Tarasov}). These two aspects grew essentially independently initially. For a detailed review of the evolution of these ideas and subsequent developments (including the Fokas method \cite{fokas1997unified,fokas2002integrable}), we refer the reader to the introduction of \cite{CCD} where a synthesis of the relationships between the various points of view was given. In Sklyanin's approach, the main ingredient is the reflection matrix $K(t;\lda)$  which is required to satisfy the (time-dependent) boundary zero curvature equation (written here in the case of the nonlinear Schr\"odinger (NLS) equation)
\begin{equation}
	\label{time_dep_boundary_ZC}
	\frac{\partial}{\partial t}K(t;\lda)=V(0,t;-\lda)\,K(t;\lda)-K(t;\lda)\,V(0,t;\lda)\,,
\end{equation}
where $V(x,t;\lda)$ is the Lax matrix for the time part of the auxiliary problem. The main idea in Habibullin's approach is to use a B\"acklund transformation, realised by a B\"acklund-Darboux matrix $L(x,t;\lda)$, to obtain solutions on the half-line from solutions on the full-line with special symmetries on their scattering data. As exhibited for instance in \cite{BT1, Tarasov, CZ,  CCAL,CCD}, the connection between the two points of view boils down to
\begin{equation}
K(t;\lda)=L(0,t;\lda)\,.
 \end{equation}

The initial-boundary value problems obtained by Sklyanin's criterion, in the case of time-independent reflection matrices $K(\lda)$, fall into the class of special initial-boundary value problems called linearizable in the Fokas method. They are amenable to Habibullin's {\it nonlinear mirror image method} (terminology adapted from \cite{biondini2009solitons}) which offers the advantage of producing multi-soliton solutions explicitly. Of course the Fokas method is powerful in the sense that it allows one to also consider boundary conditions which break integrability at the boundary. In this respect, we note that a phenomenon of emission of solitons from the boundary was analysed in \cite{FI} using the early ideas of the Fokas method\footnote{We are grateful to G. Biondini for pointing out this reference to us.}. In that context, the production of solitons by the boundary was identified indirectly using long-time asymptotic analysis via a Riemann-Hilbert formulation of the solution of the focusing NLS. The latter solution was not written in closed form and did not correspond to integrable boundary conditions as opposed to the present work. To compare those results with ours within the Fokas method, the notion of time-dependent linearizable boundary conditions should be introduced and studied. This was already identified in \cite{CCD} as a necessary ingredient in the programme of unification of methods for initial-boundary value problems for integrable PDEs.

There are renewed interest and various motivations for studying integrable boundary conditions beyond the time-independent case, as illustrated by the series of works \cite{ZAMBON,ACC,CCD, Gruner2,Xia, ZC1,  Zist-half-line}. In \cite{CCD}, it was demonstrated that the nonlinear mirror image can be successfully extended to the case of time-dependent reflection matrices $K(t;\lda)$. This led to a new and exciting phenomenon of absorption or emission of solitons by the boundary. In turn, this is consistent with the intuition behind the derivation \cite{ACC} of \eqref{time_dep_boundary_ZC} in the Hamiltonian picture for a {\it non-diagonal} reflection matrix which is the signal that ``particles'' can be absorbed or emitted by the boundary.

The first objective of this paper is to develop the investigation of time-dependent reflection matrices beyond the simplest case of \cite{ZAMBON} (see also \cite{CCD,Gruner2,Xia}) and to analyze the corresponding absorption/emission processes that can take place for solitons and higher-order solitons on the half-line. This leads to an interesting phenomenon that the usual charges such as the number of solitons are no longer conserved. The particular model we consider is the focusing NLS on the half-line with fast-decaying behavior at infinity. Instead of using the time-dependent version of the nonlinear mirror image method introduced in \cite{CCAL,CCD}, we adopt the recent method of \cite{Zist-half-line} which allows for the reconstruction of the solution $q(x,t)$ of the half-line problem {\it directly} from Sklyanin's double-row monodromy matrix. This bypasses the need for the construction of an adequate B\"acklund-Darboux matrix and eliminates the step involving the mapping to the full-line problem. The latter point is similar to the point of view taken in the Fokas method. However, the method differs in that the scattering data required for the reconstruction of $q(x,t)$ is drawn directly from the double-row monodromy matrix rather than being built from two separate sets of scattering data which are then required to satisfy a global relation, see \eg \cite{fokas2002integrable}. We note that the results on the symmetries of the scattering data (\cf \eqref{sym_A} below) were already obtained \cite{HH2} using Habibullin's B\"acklund transformation method\footnote{We thank the reviewer for bringing this reference to our attention.}. In this sense, our findings within the construction of the double-row monodromy matrix represent an alternative derivation of those symmetries. Further comparisons between \cite{HH2} and our results will be made in Remark \ref{rmk:cprs}.

The second objective of this paper is to lay the ground for a possible quantization of integrable PDEs with time-dependent boundary conditions and to make contact with older results concerning the quantum NLS with integrable boundary conditions \cite{GLM1,GLM2,MRS}. Specifically, we present the Hamiltonian description of our model with time-dependent boundary conditions and derive explicitly for the first time the classical version of the boundary (Zamolodchikov-Faddeev) algebra introduced in \cite{bound-alg}. The latter was used to quantize NLS with Robin boundary conditions in \cite{GLM1,GLM2} and to study the quantum symmetries of this model \cite{MRS} (in the vector-valued case). We derive our results in the absence of any discrete spectrum, as this is the only regime for which (second) quantization is known, to the best of our knowledge.  

The paper is organized as follows. In Section \ref{LP}, we review the Lax pair formalism for integrable boundary conditions. We also present a general result (Prop. \ref{prop:Kexpansion}) about the structure of the boundary conditions we are interested in and their characterization in terms of a time-dependent reflection matrix $K(t;\lda)$. In Section \ref{solutions}, we explain how this reflection matrix enters Sklyanin's double-row monodromy matrix $\Gamma(t;\lda)$ and extend the results of \cite{Zist-half-line} which allow one to reconstruct (soliton) solutions of the NLS equation with time-dependent boundary conditions directly from $\Gamma(t;\lda)$. The results are then illustrated by a variety of explicit examples where phenomena of absorption and/or emission of solitons at the boundary occur. Section \ref{Ham} shifts the emphasis to the Hamiltonian formulation of the model at hand. Our main motivation for this section is to lay the ground for a possible quantization. The main result in this section is the explicit derivation of a classical version of the boundary algebra introduced in \cite{bound-alg}. Finally, in Section \ref{Ccl} are gathered our conclusions which are put into perspective and related to open questions.

\section{Lax pair formalism 
  and integrable boundary conditions}\label{LP}
This section first recalls the bulk and boundary Lax pair formalism, also known as the bulk and boundary zero curvature equation respectively, for the focusing NLS equation on the half-line. The boundary zero curvature equation enables us to derive a class of integrable time-dependent boundary conditions. 
\subsection{ Lax pair formalism for NLS on the half-line}
We recall that the focusing non-linear Schr\"odinger equation for the complex field $q:=q(x,t)$ on the positive half-line reads
\begin{equation}\label{eq:NLS}
  iq_t + q_{xx} + 2 |q|^2q = 0\,,   \quad \text{for}\quad 
 x\geq 0 \,, 
\end{equation}
where the subscripts $x$ and $t$ denote partial derivative with respect to $x$ and $t$ respectively. We assume that the NLS field $q(x,t)$ and its derivatives vanish rapidly as $x\to \infty$. In other words, the {\it vanishing boundary conditions} are imposed at infinity. We shall impose boundary conditions at $x=0$ (see below) such that there still exists a description in terms of the Lax pair. 

First of all, let us  summarize the well-known results about the Lax pair formalism.
In this context, the bulk equation \eqref{eq:NLS} is rewritten equivalently as follows
\begin{eqnarray} \label{eq:c0}
 U_t(x,t;\lambda)-V_x(x,t;\lambda)+[ U(x,t;\lambda),V_x(x,t;\lambda)]=0,\
\end{eqnarray}
known as the (bulk) zero curvature equation,  where 
\begin{equation}\label{eq:UVee}
  U(x,t;\lambda) = -i \lambda \sigma_3 +Q(x,t) \,,\quad V(x,t;\lambda) = \big(-2i\lambda^2+|q(x,t)|^2\big)\,\sigma_3+2\lambda \, Q(x,t) -iQ_x(x,t)\,\sigma_3\,,
\end{equation}
with
\begin{equation}
  \label{eq:laxp2}
   Q(x,t)=\bma 0 & q(x,t) \\ -q^*(x,t) & 0 \ema\,,\quad    \sigma_3=\bma 1 & 0 \\ 0& -1 \ema\,. 
\end{equation}
We call $U$ and $V$ respectively the space-part and time-part of the Lax pair $(U,V)$.
They are traceless matrices, and obey the  involution relations
\begin{equation}\label{eq:inv1}
  U^*(x,t;\lambda^*) = \sigma_2\,U(x,t;\lambda)\, \sigma_2  \,,\quad V^*(x,t;{\lambda}^*) = \sigma_2\,V(x,t;\lambda)\, \sigma_2\,,\quad \sigma_2=\bma 0 & -i \\ i& 0 \ema\,,
\end{equation}
where the superscript  ${}^*$ denotes the complex conjugate.
 Due to the vanishing boundary conditions imposed as $x\to \infty$, one has
\begin{equation}
  \lim_{x\to+ \infty} U(x,t;\lambda) = -i\lambda\sigma_3\,,\quad  \lim_{x\to +\infty} V(x,t;\lambda) = -2i\lambda^2\sigma_3\,.
\end{equation}

Similarly to the bulk equation, we require that the boundary conditions at $x=0$ can be written in the form of a boundary zero curvature equation
\begin{equation}\label{eq:c1}
 K_t (t;\lambda) = V(0,t;-\lambda)K(t;\lambda)-K(t;\lambda)V(0,t;\lambda)\,, 
 \end{equation}
for some $2\times 2$ time-dependent matrix $K$, which is called reflection matrix, or simply $K$-matrix. The $K$-matrix is assumed to be nondegenerate for generic $\lambda$.  The strategy is to find an appropriate $K$-matrix to determine the boundary conditions.
Assuming the local existence of a $2 \times 2$ matrix $\Psi(x,t;\lambda)$ satisfying
\begin{subequations}
  \label{eq:zccbb}
\begin{align}
\label{eq:laxeqs} &\Psi_x(x,t;\lambda)=U(x,t;\lambda) \Psi(x,t;\lambda)\ , \quad\Psi_t(x,t;\lambda)=V(x,t;\lambda) \Psi(x,t;\lambda)\ ,\\
 & \Psi(0,t;\lambda)=K(t;\lambda) \Psi(0,t;-\lambda)\,,
\end{align}  
\end{subequations}
  \eqref{eq:c0} and \eqref{eq:c1} can be seen as the consistency conditions for this system.

\subsection{Time-dependent boundary conditions \label{sec:Km}}

Before deriving explicit forms of the $K$-matrices, let us recall some  properties of the reflection matrices  \cf \cite{Zist-half-line}.
It follows from the fact the time-part Lax matrix $V$ is traceless that $\det K(t;\lambda)$ is time-independent and is denoted by $\kappa(\lambda)$, \ie
\begin{equation} \det K(t;\lambda) =\kappa(\lambda) \,.\end{equation}
Since equation \eqref{eq:c1} is linear in $K$, it does not fix its normalization. 
For later convenience, it is fixed in the following way:
\begin{equation}
\label{eq:ks}    K^{-1}(t;\lambda) = K(t;-\lambda)\,,\quad \kappa(-\lambda)K(t;\lambda)=\sigma_2{K}^*(t;{\lambda}^*)\sigma_2\,,
  \end{equation}  
which is equivalent to requiring  
\begin{equation}\label{eq:ff1}
    \kappa(\lambda)\kappa(-\lambda) =1\,,\quad {\kappa}^*({\lambda}^*) =\kappa(-\lambda)\,.
\end{equation}

If we assume that the asymptotic of $K$ is fixed as follows
\begin{equation}\label{eq:limkm}
      \lim_{\lambda \to \infty} K(t;\lambda) = \bma 1 &0 \\ 0& \epsilon \ema\,,
\end{equation}
and that $\kappa$ is a rational function in $\lambda$, then, due to \eqref{eq:ff1}, it can be expressed in the form
\begin{equation}\label{eq:normk1}
\kappa(\lambda) =  \epsilon \prod_{j=1}^{\cN}\left[\frac{\lambda-\beta_{j}}{\lambda-\beta_{j}^{*}}\frac{\lambda+\beta_{j}^{*}}{\lambda+\beta_{j}}\right]\prod_{\ell=1}^{\cM}\left[\frac{\lambda-i\alpha_\ell}{\lambda+i\alpha_\ell}\right]\,,\quad \epsilon =\pm 1\,,
\end{equation}
where $\beta_j$ is a complex number with a non-vanishing real part, 
and $\alpha_\ell$ is a non-vanishing real number. 
Then, if the determinant of $K$ is fixed as \eqref{eq:normk1},
 the  $K$-matrix can be uniquely determined by the relation \eqref{eq:c1}.

Here, we focus on the solutions $K(t;\lambda)$ of equation \eqref{eq:c1} having a determinant given by
\eqref{eq:normk1} with $\cM=0$ and $\epsilon=1$. 
Without loss of generality, we restrict ourselves to the case where the imaginary part of $\beta_i$ is negative and the real part is nonzero. This yields a class of time-dependent $K$-matrices accompanied by a class of time-dependent integrable boundary conditions. 

\begin{proposition}\label{prop:Kexpansion}
The solution to equation \eqref{eq:c1} satisfying \eqref{eq:limkm} and with a determinant 
\begin{equation}\label{eq:kappa}
 \kappa(\lambda)=  \prod_{j=1}^{\cN}\left[\frac{\lambda-\beta_{j}}{\lambda-\beta_{j}^{*}}\frac{\lambda+\beta_{j}^{*}}{\lambda+\beta_{j}}\right]\,,
\end{equation}
takes the form 
\begin{equation}\label{eq:Kex}
      K(t;\lambda)= \prod_{j=1}^{\cN}\left[\frac{1}{\lambda-\beta_{j}^{*}}   \frac{1}{\lambda+\beta_{j}}\right]
      \left(     \sum_{j=0}^{\cN}  \lambda^{2j} f_{2j}(t)  1\!\!1 +\  i \sum_{j=0}^{\cN-1}  \lambda^{2j+1} \begin{pmatrix}
              f_{2j+1}(t)  &   g_{2j+1}(t)\\
              g^*_{2j+1}(t)  & - f_{2j+1}(t) 
                \end{pmatrix}\right)\,, 
\end{equation}
where $1\!\!1$ is the identity matrix, $f_j(t)$ are real-valued functions and $g_j(t)$ are complex-valued functions, with $f_{2\cN}(t)=1$.  
Then, the functions $g_j(t)$ can be expressed as 
\begin{equation}\label{eq:g}
g_{2j-1}(t) = D_t^{\cN-j}q -\sum_{k=0}^{\cN-j-1} D_t^k\Big(-qf_{2(j+k)}(t) +\frac12 q_xf_{2(j+k)+1}(t)\Big),\quad 1\leq j\leq \cN\,
\end{equation}
where $D_t : =\frac{i}4\,\frac{d}{dt} +\frac12|q|^2$, and $q$ stands for $q(0,t)$. The functions $f_j(t)$ obey,  for $0\leq p \leq 2\cN$,
\begin{equation}\label{eq:detc}
 \sum_{0\leq k,\ell\leq 2\cN \atop k+\ell=2p} \Big( f_k(t) f_\ell(t)+g_k^*(t) g_\ell(t) \Big)=
 (-1)^p\ \tau_{2\cN-p}\big(  \beta_1^2,\dots,\beta_\cN^2, (\beta^*_1)^2,\dots,(\beta^*_\cN)^2 \big)\,,
\end{equation}
where  by convention, $g_{2j}(t)=g^*_{2j}(t)=0$, $0\leq j \leq \cN$.
$\tau_n$ is the $n^{\text{th}}$  elementary symmetric polynomial.

The boundary relation associated to a $K$-matrix  of the form \eqref{eq:Kex} reads
\begin{equation}\label{eq:bord}
 D_t^{\cN}q  = \sum_{k=0}^{\cN-1} D_t^k \Big(-qf_{2k}(t) +\frac12q_xf_{2k+1}(t)\Big)\,,
\end{equation}
where the functions $f_k(t)$ are subject to 
the constraints \eqref{eq:detc} with $g_k^*(t)$ and $g_\ell(t)$ replaced by their expressions \eqref{eq:g}.
\end{proposition}
\prf
The condition \eqref{eq:limkm} with $\epsilon=1$ implies that $f_{2\cN}(t)=1$.
Plugging the expansion \eqref{eq:Kex} into equation \eqref{eq:c1} we get, for $0\leq j \leq \cN-1$,
\begin{subequations}
 \begin{align}
    & i\frac{d}{dt}f_{2j}(t)=  2 q g^*_{2j-1}(t) - 2 q^* g_{2j-1}(t) \,, \quad
   i\frac{d}{dt}f_{2j+1}(t)=  q_x^* g_{2j+1}(t) -q_x g^*_{2j+1}(t)  \,, \label{eq:aj}\\
&  \frac{i}{2}\frac{d}{dt}g_{2j+1}(t)= 2g_{2j-1}(t)  -  |q|^2g_{2j+1}(t) -2  q f_{2j}(t) + q_x f_{2j+1}(t) \,, \label{eq:bj}\\
 &g_{2\cN-1}(t)= q\,, \label{eq:gN}
 \end{align}  
\end{subequations}
where by convention $g_{-1}(t):=0$. In particular, one gets $\frac{d}{dt}f_0(t)=0$. These relations are equivalent to \eqref{eq:c1} once the expansion of $K$ is assumed.
Equations \eqref{eq:aj} are compatible with the reality of the functions $f_j(t)$ and equations for $g_{2j+1}^*(t)$ are deduced from \eqref{eq:bj} by complex conjugation.

Comparing the explicit form of $\kappa(\lambda)$ and a direction computation of the determinant of \eqref{eq:Kex}, one gets \eqref{eq:detc}.
It can be shown that, keeping equations \eqref{eq:bj}, the set of equations \eqref{eq:aj} can be replaced by the set \eqref{eq:detc}.

Then, starting from \eqref{eq:gN}, equations \eqref{eq:bj} for $j=\cN-1,\,\cN-2,\dots,1 $, allows us to express recursively $g_{2j-1}(t)$  in terms of  $q$, its derivative and the functions $f_k(t)$ to obtain \eqref{eq:g}.
Finally, the last equation \eqref{eq:bj} for $j=0$ provides the boundary relation. \finprf
\medskip

It can be shown, see Appendix \ref{Appendix_dressing}, that the $K$-matrix of Prop. \ref{prop:Kexpansion} has the dressed structure familiar in Sklyanin's approach \cite{SKBC}. This structure is also the main object of the construction in \cite{Zist-half-line}.

The above results provide for the NLS equation \eqref{eq:NLS} a class of integrable time-dependent boundary conditions that are completely characterized by the {\it degree} $\cN$ and the boundary parameters $\beta_j$,  $j = 1, \dots, \cN$. There is, in general, no restriction on the multiplicity of zeros (and poles) for $\kappa(\lambda)$. In other words, several $\beta_j$,  $j = 1, \dots, \cN$, could take the same value. 
The functions $f_j(t), g_j(t)$, $j = 1, \dots, 2\cN$ appearing, in the $K$-matrix can be interpreted as the some degrees of freedom  that are coupled to the NLS fields at the boundary, and the time-dependent boundary conditions are actually expressions of some nonlinear differential-algebraic system, \ie \eqref{eq:detc},  \eqref{eq:bord} and its complex conjugate.
In the following, we show explicit examples for $\cN = 1, 2$.    
  \paragraph{Example $\mathbf{\cN=1}$.} In this case, $f_0$ and $f_1(t)$ are given by $f_0^2=|\beta_1|^4$ and $f_1(t)^2+2f_0+|q|^2=-\beta_1^2-(\beta_1^*)^2$.
Since $f_0$ and $f_1(t)$ are real, the previous two relations become 
\beq
f_0=-|\beta_1|^2\mb{and} f_1(t)^2=-|q|^2-(\beta_1-\beta_1^*)^2\,.
\eeq
 Applying the procedure described above, the boundary condition becomes 
 \begin{eqnarray}\label{eq:bound-N=1}
 \frac{i}{2}\frac{d}{dt}q=-|q|^2q+2q|\beta_1|^2+q_x f_1(t)\,.
  \label{eq:bc1}
 \end{eqnarray}
We recover the time-dependent boundary conditions introduced in \cite{ZAMBON} whose solutions are studied in \cite{CCD, Gruner2, Xia}.   An equivalent form of \eqref{eq:bound-N=1}, involving the second-order derivative of $q_{xx}$ obtained by eliminating $q_t$ using the NLS equation \eqref{eq:NLS} continued at $x=0$, was obtained earlier in \cite{BT1,HH2}.

 \paragraph{Example $\mathbf{\cN=2}$.} In this case, one gets
 \begin{subequations}
 \begin{eqnarray}
    &&\frac{i}{2}\frac{d}{dt}g_1(t)=-|q|^2g_1(t)-2qf_0(t)+q_x f_1(t) \,, \label{eq:bq}\\
    &&\frac{i}{2}\frac{d}{dt}q=2g_1(t)-|q|^2q-2qf_2(t)+ q_x f_3(t)\,, \label{eq:b12}
 \end{eqnarray}
and 
\begin{eqnarray}
 && f_0^2=\tau_4\,, \qquad 2f_0f_2(t)+f_1(t)^2+|g_1(t)|^2=-\tau_3\,,  \qquad f_4=1\,,\\
 && 2f_0+2f_1(t)f_3(t)+f_2(t)^2+q g_1(t)^* +q^*g_1(t)=\tau_2\,, \quad f_3(t)^2+2f_2(t)f_4(t)+|q|^2=-\tau_1\,,
 \qquad \qquad
\end{eqnarray}
 \end{subequations}
  where the variables entering the symmetric polynomials $\tau_i$ are $\beta_1^2,\beta_2^2, (\beta^*_1)^2$ and $(\beta^*_2)^2$.
 As explained in the generic case, we can determine $g_1$ using \eqref{eq:b12} and use this expression to transform \eqref{eq:bq} to the following boundary condition
 \begin{equation}\label{eq:bound-N=2}
 \frac{1}{2}\frac{d^2 q}{dt^2}=  8qf_0 -4q_xf_1(t) +2 f_2(t)(2q|q|^2+iq_t)-f_3(t)(iq_{xt}+2q^2q^*_x)+2i|q|^2q_t-|q_x|^2q+q_x^2q^*+2q|q|^4\,.
 \end{equation}
   Note that in the process of simplification of the boundary condition, it is easier to use   \eqref{eq:aj} rather than its equivalent form \eqref{eq:detc}.
Expression  \eqref{eq:detc} is useful to get a factorization of the determinant $ \kappa(\lambda)$, \ie \eqref{eq:kappa}.

 \section{Soliton solutions of NLS on the half-line with time-dependent boundary conditions}\label{solutions}

In this section, we construct multi-soliton solutions of NLS on the half-line subject to the integrable boundary conditions given in Prop. \ref{prop:Kexpansion}. This relies on the method of \cite{Zist-half-line} which allows for reconstruction of the half-line solutions directly from the double-row monodromy matrix.  We also explore  the phenomena of absorption/emission of solitons by the boundary.  
 
 \subsection{Double-row monodromy matrix} We briefly recall the notion of double-row monodromy matrix \cite{SKBC, sklyanin1988boundary} (see also \cite{ACC} for a recent survey).   
 The scattering functions, as entries of the double-row monodromy matrix, encode the initial-boundary data, and allow us to reconstruct exact solutions of NLS on the half-line \cite{Zist-half-line}. They also play a central role in the Hamiltonian formalism of the model (see Section \ref{sec:Ham}).

 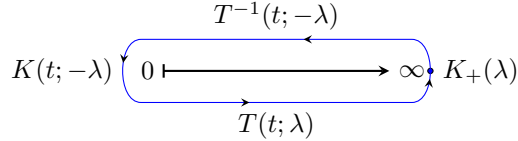
\begin{figure}[h]\centering  \begin{tikzpicture}[scale=0.6]
    \tikzstyle{nod}= [circle, inner sep=0pt, fill=black, minimum size=2pt, draw]
    \tikzstyle{nodb}= [circle, inner sep=0pt, fill=blue, minimum size=2pt, draw]
    \draw[-stealth, thick, black] (-2.5,0) -- (2.5,0);
    \draw[thick, black] (-2.5,0.15) -- (-2.5,-0.15);
    \node[left] (x0) at (-2.5,0)  {$0$};
    \node[right] (x0) at (2.5,0)  {$\infty$};
    \node[nodb] (x0) at  (3.41,0) [label=right:$K_+(\lambda)$] {};
    \node[left] (x0) at  (-3.41,0) {$K(t;-\lambda)$};
    \node[below](a) at (0,-0.7){$ T(t;\lambda) $};
    \node[above](a) at (0,0.7){$ T^{-1}(t;-\lambda) $};
  \path [draw=blue,postaction={on each segment={mid arrow=black}}]
  (-3,-0.7) to  (3,-0.7)
  (3,0.7) to   (-3,0.7)
  (3,-0.7)to[out=0,in=0]  (3,0.7)
  (-3,0.7)to[out=-180,in=-180]  (-3,-0.7)
  ;

  \end{tikzpicture}
  \caption{Double-row monodromy matrix following the path circling the semi-axis.}
  \label{fig:monomatrix}
\end{figure}
The double-row monodromy matrix $\Gamma(t;\lambda)$  can be constructed following a
path circling the positive semi-axis as depicted in Figure \ref{fig:monomatrix} by employing  the ``single-row'' transition matrix $T(t;\lambda)$ and the reflection matrices $K(t;\lambda)$ and $K_{+}(\lambda)$ (we assume $K_+(\lambda)$  to be time-independent,  this point will be discussed below after Prop. \ref{prop:Kp}) located respectively at $x = 0$ and at {\it infinity} as
\begin{equation}\label{GGGG}
  \Gamma(t;\lambda) = T(t;\lambda)\,K(t;-\lambda)\, T^{-1}(t;-\lambda)\,K_+(\lambda) \,. 
\end{equation}
The single-row transition matrix $T(t;\lambda)$ (assuming it exists) is defined as \cite{fokas2002integrable}
\begin{equation}\label{eq:TT1tm}
  T(t;\lambda) =   \lim_{L\to \infty}\left(e^{i\lambda L \sigma_3}\, T(L,0,t;\lambda)\right)  \,,
\end{equation}
with  $T (x, y,t; \lambda)$ being the fundamental solution of the space-part of the Lax equations, \ie \begin{equation}
  T (x, y,t; \lambda)=\overset{\curvearrowleft}{\exp} \int_y^x U(\xi,t;\lambda) d\xi \ ,\quad  T (x, x,t; \lambda)=1\!\!1\,,
 \end{equation}
 where $\overset{\curvearrowleft}{\exp}$ denotes the path-ordered exponential. The reflection matrix  $K(t;\lambda)$  is supposed to be one of the solutions provided in Prop. \ref{prop:Kexpansion}. Therefore, the  time-dependent boundary conditions associated with $K(t;\lambda)$ are imposed  at $x =0$.   The form of the reflection matrix $K_+(\lambda)$  will be characterized in Prop. \ref{prop:Kp} below in order to make the half-line NLS model integrable. 

 Note that, for the above definition of $ \Gamma (t; \lambda)$,  the choice of the starting point of the path of integration as well as the orientation of the path, is irrelevant to the spectral properties of $\Gamma(t; \lambda)$. In our case, we fix the starting point at infinity and make the path anti-clockwise (see Figure \ref{fig:monomatrix}). We also assume that the NLS field $q(x,t)$ belongs to the functional space of Schwartz-type for $x\geq 0$ at a fixed time $t$ so that the single-row transition matrix  $T(t;\lambda)$ \eqref{eq:TT1tm} is well-defined.
 
 \begin{proposition}\label{prop:Kp}
Assume the single-row transition matrix  $T(t;\lambda)$ \eqref{eq:TT1tm} is well-defined, and let $K(t;\lambda)$  be one of the solutions provided in Prop. \ref{prop:Kexpansion} with the normalization
   \begin{equation}
\det K(t;\lambda)=     \kappa(\lambda) \,, 
   \end{equation}
 where $\kappa(\lambda)$ is given in \eqref{eq:kappa}. If  $K_+(\lambda)$ is a time-independent matrix such that
\begin{subequations}\label{eq:propKp}
 \begin{align}
   \label{eq:Kprop} [\sigma_3, K_+(\lambda)] &= 0\,, \\
                  \label{eq:Kprop1}                            \det K_+(\lambda)&  = \kappa(\lambda)\,,\\
 \label{eq:assumK+}
     \kappa(\lambda)\sigma_2{K}_+^*({\lambda}^*)\sigma_2 & = K_{+}(\lambda)\,, \\ \label{eq:assumK2} \lim_{\lambda \to \infty} K_+(\lambda)& = 1\!\!1\,, 
  \end{align}
  then the inverse part of the ISM applied to $\Gamma(t;\lambda)$ leads to solutions of NLS on the half-line subject to the time-dependent boundary conditions at $x=0$ associated with $K_(t;\lambda)$.
\end{subequations}
 \end{proposition}

 In other words, the set of constraints \eqref{eq:propKp} provides a sufficient condition for the half-line NLS model to be integrable by means of the ISM. Proof of this statement can be found in   \cite[Theorem $4.5$]{Zist-half-line}. Actually, the integrability can already be seen through the properties of the double-row monodromy matrix, such as the Lax formulation \eqref{eq:LF} and  the unimodularity listed below. The ISM requires extra analytic properties of the scattering system associated with $\Gamma(t;\lambda)$, and the inverse part of the ISM can be formulated as a Riemann-Hilbert problem. Also note that the assumption that $K_+(\lambda)$ is a time-independent matrix can be regarded as a consequence of  $K_+(\lambda)$ also satisfying the boundary zero curvature equation \eqref{eq:c1}  evaluated as $x\to \infty$.  

 Let us comment on some direct consequences of the above characterizations of the double-row monodromy matrix $\Gamma(t;\lambda)$.   \begin{itemize}
 \item  $\Gamma(t;\lambda)$ satisfies the following Lax formulation
   \begin{equation}
     \label{eq:LF}\frac{d}{dt}\Gamma(t;\lambda) = \lim_{x\to\infty}[ V(x,t;\lambda), \Gamma(t;\lambda) ] = -2i\lambda^2 [\sigma_3, \Gamma(t;\lambda)]\,. 
   \end{equation}
This can be derived using \eqref{eq:c1} and \eqref{eq:Kprop}, and the fact that the  vanishing boundary conditions as $x\to \infty $ are imposed.
 \item  $ \Gamma(t;\lambda)$ is unimodular, \ie   $ \det\Gamma(t;\lambda) = 1$, and satisfies the involution relation
   \begin{equation}\label{eq:invg}
       \Gamma^*(t;\lambda^*) = \sigma_2\,\Gamma(t;\lambda)\, \sigma_2\,. 
   \end{equation}
   \item One can express $\Gamma(t;\lambda)$ as
   \begin{equation}\label{eq:G-matrix}
      \Gamma(t;\lambda) = \bma A(\lambda  ) &  -e^{-4i\lambda^2 t}B^*(\lambda^*) \\  e^{4i\lambda^2 t}B(\lambda)  & A^*(\lambda^*)\ema\,.   
    \end{equation}
The scattering functions $A(\lambda), B(\lambda)$ are well-defined for $\lambda\in \RR$, and  $A(\lambda)$ ({\it resp}. $A^*(\lambda^*)$)  can be analytically extended off the real axis to the upper   ({\it resp}.   lower) half complex plane\footnote{We assume the boundary parameters $\beta_j$, $j=1,\dots,\cN$, have negative imaginary parts.}, and behaves asymptotically as 
    \begin{equation}
      \lim_{\lambda\to \infty}A(\lambda) = 1 + O(1/\lambda)\,. 
    \end{equation}
    \item $\Gamma(t;\lambda)$  satisfies a second involution relation    \begin{equation}\label{eq:inv222}
      \Gamma^{-1}(t;-\lambda) =  K^{-1}_+(-\lambda) \Gamma(t;\lambda)K^{-1}_+(\lambda)\,. 
    \end{equation}

 \end{itemize}

Recall the form of $\kappa(\lambda)$ given in \eqref{eq:kappa}. It follows from the constraints \eqref{eq:propKp} that the reflection matrix $K_+(\lambda)$ should be put in the form
  \begin{equation}
\label{K_split}
  K_+(\lambda) = \bma  \varphi(\lambda) & 0 \\ 0 &  \kappa(\lambda)/\varphi(\lambda)  \ema \,,
\end{equation}
where  $\varphi(\lambda)$ is given by  
\begin{equation}\label{eqLphi}
  \varphi(\lambda) = 
  \prod_{j \in \cJ}  \left[\frac{\lambda-{\beta}_{j}}{\lambda-{\beta}_{j}^{*}}\right] 
  \prod_{k \in \cK}  \left[\frac{\lambda+{\beta}_{k}^{*}}{\lambda+{\beta}_{k}}\right]\,,
\end{equation}
with  $\cJ$, $\cK$ being disjoint subsets of $\{1,\dots, \cN\}$. 

\begin{rmk}
  We stress that the form of $K_+(\lambda)$ given in \eqref{K_split} along with the involution relation \eqref{eq:inv222} represents a crucial generalization to the results given in   \cite{Zist-half-line}. Such form of $K_+(\lambda)$ leads to certain symmetries of the scattering functions which characterize initial-boundary data for the half-line problem under consideration. In particular, novel types of absorption/emission of solitons by the boundary can be constructed. 
\end{rmk}

Using the form  of $K_+(\lambda)$ given in \eqref{K_split}, it follows from \eqref{eq:inv222} that  $A(\lambda), B(\lambda)$,  for $\lambda\in \RR$,   satisfy  the following symmetric relations 
  \begin{equation}\label{sym_A}
 A(-\lambda)=g_A(\lambda)
  A^*(\lambda^*)\,,\quad B(-\lambda) =g_B(\lambda)
  B(\lambda)\,,
\end{equation}
 where
\begin{subequations}\label{eq:GGG}
\begin{align}
 & g_A(\lambda)=\varphi(\lambda)\varphi(-\lambda)=
 \prod_{j \in \cJ}  \left[\frac{\lambda^2-{\beta}_{j}^2}{\lambda^2-({\beta}_{j}^{*})^2}\right] 
  \prod_{k \in \cK}  \left[\frac{\lambda^2-({\beta}_{k}^{*})^2}{\lambda^2-({\beta}_{k})^2}\right],
  \label{eq:gA}\\
&  g_B(\lambda)=-\frac{\kappa(\lambda)\varphi(-\lambda)}{\varphi(\lda)} = -
    \prod_{i \in {\cI}}  \left[\frac{\lambda-{\beta}_{i}}{\lambda-{\beta}_{i}^{*}} \frac{\lambda+{\beta}_{i}^{*}}{\lambda+{\beta}_{i}}\right],
 \label{eq:gB}
\end{align}
\end{subequations}
 with  $\cJ$, $\cK$ being disjoint subsets of $\{1,\dots, \cN\}$ and $ {\cI} = \{1,\dots, \cN\} \setminus (\cJ \cup \cK) $. In particular, $B(\lambda)$ obeying the above symmetric relation characterizes the so-called {\it continuous scattering data} of the half-line NLS model under consideration.    
  
 \begin{rmk}\label{rmk:cprs} It should be noted that similar results as \eqref{sym_A} were already obtained by Habibullin (see (1.4) in \cite{HH2}). It was based on a symmetric extension of the monodromy matrix which coincides with the involution relation of the double-row monodromy matrix \eqref{eq:inv222}. The associated boundary conditions were derived using a B\"acklund transformation by exploring the space reversal symmetry of the NLS equation. In the present paper, \eqref{eq:inv222} as well as the boundary conditions are derived as consequences of the construction of  $\Gamma(t;\lambda)$. These two approaches reflect the two paths for the same problem initiated respectively by Habibullin and Sklyanin, and are eventually connected as mentioned in Introduction. However, we stress that a detailed analysis of the solution content which derives from the symmetry on the scattering data, in particular the possibility of emission/absorption of solitons at the boundary and the consequences for conserved quantities, as presented in the next (sub)sections, were not discussed in \cite{HH2}.   
 \end{rmk}

\subsection{Absorption and/or emission   of solitons by the boundary}
Let us first the recall the multi-soliton solutions of NLS  \cite{FT}. The $N$-soliton solutions of NLS on the full-line are expressions of $2N$ nonzero complex parameters  $\{\lambda_j; c_j\} $, $ j =1, \dots, N$, known as the {\it discrete scattering data}. Here,  $\lambda_j$ denotes the discrete spectrum having positive imaginary part, and $c_j$ is the {\it norming constant} associated with $\lambda_j$. We also assume that the discrete spectrum are distinct from each other. Then,   the $N$-soliton solutions of NLS for $-\infty<x<\infty$ can be put in the form 
\begin{equation}
    \label{expression_u}
    q(x,t)= \frac{-2i  }{ \det  M}\begin{vmatrix} 0& \begin{matrix}   1 & \cdots & 1  \end{matrix}  \\ \begin{matrix}   \gamma_{1} \\ \vdots \\ \gamma_{N} \end{matrix} &  M       \end{vmatrix} \,, 
    \end{equation}
    where $\gamma_j :=  c_j e^{-2{ i \lambda_j}( x+2\lambda_j t) }$, and  $M$ is a $N \times N$ matrix with  components     $ M_{j\ell}=\frac{{\gamma}_j{\gamma}^*_\ell+1}{\lambda_j-\lambda^*_\ell}$. This formula will be employed later in the  construction of multi-soliton solutions of NLS on the half-line. Note that the large-time asymptotics of the $N$-soliton solutions leads to $N$  independent solitons. Let $ \mu_j$, $\nu_j $ (assume  $\nu_j>0$) be respectively the real and imaginary parts of $\lambda_j$, \ie $ \lambda_j = \mu_j+i\nu_j$, then the amplitude and velocity of each asymptotic soliton associated with $\lambda_j$ are respectively $2\nu_j  $ and  $-4 \mu_j$.
 One could also consider {\it higher-order} (or {\it multi-pole}) soliton solutions by taking several discrete eigenvalues to have the same value. Then, the multi-pole soliton solutions can be obtained using certain limiting process of the above formula.  A special type of double-pole soliton fully absorbed and/or emitted by the boundary on the half-line will be constructed below. 
 Take $\kappa(\lambda)$ in the form \eqref{eq:kappa}, and without loss of generality, assume  all the boundary parameters $\beta_j$, $i =1,\dots, \cN$, are located in the fourth quadrant of the complex plane, \ie having positive real parts and negative imaginary parts. Then, as a consequence of the symmetric relation \eqref{sym_A},  the scattering function $A(\lambda)$ in the pure-soliton (or reflectionless) case can be expressed as   \begin{equation}
\label{expression_A}
A(\lda)=\prod_{\ell=1}^N\frac{\lambda-{\mu{}}_{\ell}}{\lambda-{\mu{}}_{\ell}^*}\frac{\lambda-\widetilde{\mu{}}_{\ell}}{\lambda-\widetilde{\mu{}}^*_{\ell}}\prod_{j \in \cJ}  \frac{\lambda+{\beta}_{j}}{\lambda+{\beta}_{j}^*}
\prod_{k \in \cK}  \frac{\lambda-{\beta}_{k}^*}{\lambda-{\beta}_{k}} = \prod_{\ell=1}^N\frac{\lambda-{\mu{}}_{\ell}}{\lambda-{\mu{}}_{\ell}^*}\frac{\lambda-\widetilde{\mu{}}_{\ell}}{\lambda-\widetilde{\mu{}}^*_{\ell}} \,\varphi(-\lambda)\,,
\end{equation}
where  $\varphi(\lambda)$ is given in \eqref{eqLphi} and 
\begin{equation}\label{eq:muinv}
   \widetilde{\mu}_\ell  =  - \mu^*_\ell\,.
\end{equation}
This form extracts the discrete spectrum that are zeros of $A(\lambda)$. A pair $\mu_\ell, \widetilde{\mu}_\ell$ of the spectrum, together with the paired norming constants $c_\ell, \widetilde{c}_\ell$ (see \eqref{sym_norming}), generates two solitons on the full-line with opposite velocities. This corresponds to one soliton on the half-line reflected by the boundary. The new feature here is the combinatorial aspect of relating zeros of $A(\lambda)$ with the boundary parameters. This leads to the phenomena of  absorption and/or emission of one or several solitons by the boundary, which is the content of  the following statement.

\begin{proposition}\label{prop:main}
  Take $\kappa(\lambda)$ in the form \eqref{eq:kappa} and $g_B(\lambda) $ in \eqref{eq:gB}, and assume the boundary parameters $\beta_j$, $j=1, \dots, \cN$, are located in the fourth quadrant of the complex plane. Then,  the multi-soliton solutions of NLS on the half-line subject to the boundary conditions associated with  $\kappa(\lambda)$ can be constructed using formula \eqref{expression_u}, restricting it to $x\geq 0$, with the discrete scattering data
  \begin{equation}\label{eq:refdisdata}\{\mu_\ell, \widetilde{\mu}_\ell ;  c_\ell, \widetilde{c}_\ell\}_{\ell = 1, \dots, N} \cup \{-\beta_j; \fc_j\}_{j\in \cJ} \cup \{\beta_k^*; \fc_k \}_{k\in \cK}\,,  \end{equation}
where the discrete spectrum $\mu_\ell,\widetilde{\mu}_\ell, -\beta_j, \beta_k^*$ ($\ell=1,\dots,N$,  $j\in \cJ$, $k\in \cK$) are distinct parameters, and $\widetilde{\mu}_\ell,  \widetilde{c}_\ell$ are paired with 
  ${\mu}_\ell, {c}_\ell$ as \bea\label{sym_norming}
\widetilde{\mu}_\ell = - \mu^*_\ell\,,\quad   c_\ell \widetilde{c}^*_\ell=g_B(\mu_\ell)\,. 
\eea
The norming constants $\fc_j$ and $\fc_k$ associated with $-\beta_j$ and $\beta_k^*$ are nonzero and can be freely chosen. Let $|\cJ|$ and $|\cK|$ be the cardinality of $\cJ$ and $\cK$. Then, the so-constructed solutions correspond to $(N+|\cJ|+|\cK|)$-soliton solutions on the half-line with $N$ solitons reflected,  $|\cJ|$ solitons emitted and  $|\cK|$ solitons absorbed.  
\end{proposition}

Note that \eqref{expression_u} is also used here as the {\it reconstruction formula} for half-line soliton solutions but combined with the extra symmetries of the scattering data. This is due to the fact that the inverse part of the ISM for the half-line problem (formulated as a Riemann-Hilbert problem) shares a lot of similarities with that of a full-line problem. As a consequence, soliton solutions can be reconstructed using certain Darboux-dressing procedure in complete analogy with a full-line problem. Details of the Riemann-Hilbert formulation of the half-line problem as well as the dressing procedure  can be found in \cite{Zist-half-line}. The  sketch of the proof for the symmetries of the scatteing data will be given in Appendix \ref{app:prf1}.  This implies the usual charges such as the number of solitons are no longer conserved for the half-line problem subject to the integrable time-dependent boundary conditions. However an infinite set of modified conserved quantities will be provided in the next Section by taking the boundary into account.
 
As a particular example, we provide now details of two-soliton solutions of NLS on the half-line with $\cN=2$   (see  \cite{CCD} for the case $\cN=1$). In this case, one has
\begin{equation}
  \kappa(\lambda) =  \left(\frac{\lambda-\beta_{1}}{\lambda-\beta_{1}^{*}}\frac{\lambda+\beta_{1}^{*}}{\lambda+\beta_{1}}\right) \left(\frac{\lambda-\beta_{2}}{\lambda-\beta_{2}^{*}}\frac{\lambda+\beta_{2}^{*}}{\lambda+\beta_{2}}\right)\,, 
\end{equation}
and $ N + |\cJ|+|\cK| = 2$.  
\begin{description}
\item[Case I: $\beta_1\neq \beta_2$.]   We have the following choices for  the function $\varphi(\lambda)$ (note that the zeros of $A(\lambda)$ induced by the boundary parameters are determined by $\varphi(-\lambda)$, \cf \eqref{expression_A}). \begin{itemize}
  \item For $\varphi(\lambda)=1$,  $|\cJ|=| \cK| = 0$, and $N=2$.  We recover the usual reflected soliton solutions where two-soliton solutions are reflected by the boundary.

   \item For $\varphi(\lambda) =   \frac{\lambda+\beta_{k}^*}{\lambda+\beta_{k}}$, $k=1,2$, $|\cJ|=0, |\cK|=1$, or for $\varphi(\lambda)=\frac{\lambda-\beta_{j}}{\lambda-\beta_{j}^{*}}$, $j=1,2$, $|\cJ|=1, |\cK|=0$. In both cases, $N=1$,  we can obtain two-soliton solutions where one soliton is either absorbed ($|\cK|=1$) or emitted ($|\cJ|=1$), and the other one is reflected. This is similar to the case with $\cN=1$ treated in \cite{CCD}. 
   
     \item 
   For $\varphi(\lambda)=\frac{\lambda-\beta_{1}}{\lambda-\beta_{1}^{*}} \frac{\lambda-\beta_{2}}{\lambda-\beta_{2}^{*}}$, $|\cJ| = 2, |\cK| = 0, N =0$. We get solutions where two solitons with different speed are emitted.
This is illustrated in Figure \ref{plot_2soliton_absorbedb}, where  the contour of a particular solution is displayed.
   Similarly, for  $\varphi(\lambda))=
   \frac{\lambda+\beta^*_{1}}{\lambda+\beta_{1}}\frac{\lambda+\beta_{2}^*}{\lambda+\beta_{2}}$,  $|\cJ| = 0, |\cK| = 2, N =0$, one gets two solitons absorbed by the boundary.
\item For 
     $\varphi(\lambda)=
   \frac{\lambda-\beta_{1}}{\lambda-\beta^*_{1}}\frac{\lambda+\beta_{2}^*}{\lambda+\beta_{2}}$ (or $\varphi(\lambda)=\frac{\lambda+\beta^*_{1}}{\lambda+\beta_{1}}\frac{\lambda-\beta_{2}}{\lambda-\beta_{2}^{*}}$), $|\cJ| = |\cK| =1$. This provides
    two-soliton  solutions where one soliton is emitted and the other one is absorbed (see Fig \ref{plot_2soliton_abs-em}).

   \end{itemize}
   \begin{figure}[htp]
    \centering
    \includegraphics[width=6cm]{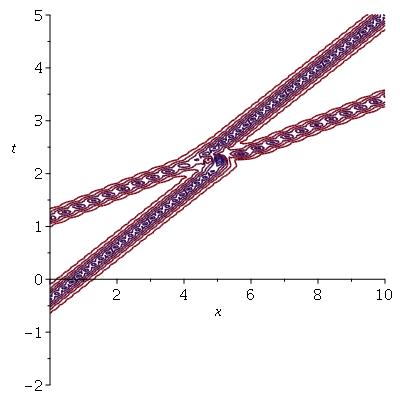}
    \hspace{2cm}
    \includegraphics[width=6cm]{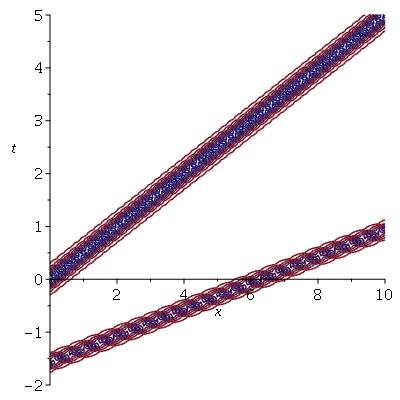}
    \caption{$2$D-contour plots of $|q(x,t)|$ corresponding to two emitted solitons with time-dependent boundary conditions of $\cN=2$. One has $\beta_1=1-2i, \beta_2=(1-5i)/2$. The scattering data are $\{-\beta_1, -\beta_2; \fc_1= e^{20},\fc_2=5 \}$ for the left figure, and 
$\{-\beta_1, -\beta_2; \fc_1= e^{-20},\fc_2=5 \}$ for the right figure. }
    \label{plot_2soliton_absorbedb}
\end{figure}

   \begin{figure}[htp]
    \centering
    \includegraphics[width=6cm]{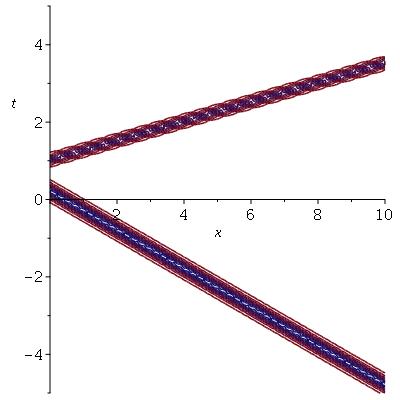}
    \hspace{2cm}
    \includegraphics[width=6cm]{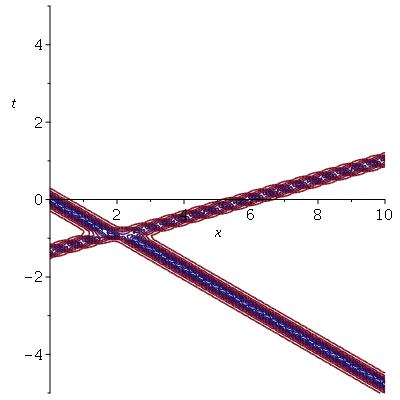}
    \caption{$2$D-contour plots of $|q(x,t)|$ corresponding to two solitons (one emitted and one absorbed) with time-dependent boundary conditions of $\cN=2$. One has $\beta_1=1-2i, \beta_2=(1-5i)/2$. The scattering data are  $\{-\beta_1, \beta^*_2; \fc_1= e^{20},\fc_2=5 \}$ for the left figure, and 
$\{-\beta_1, \beta^*_2; \fc_1= e^{-20},\fc_2=5 \}$ for the right figure.}    \label{plot_2soliton_abs-em}
\end{figure}

\item[Case II: $\beta_2 = \beta_1$]\
  Here, we explore another interesting phenomena by setting both the boundary parameters to be the same. Let $\varphi(\lambda) =(\frac{\lambda+\beta^*_{1}}{\lambda+\beta_{1}})^2$,  $|\cJ| = 0, |\cK| = 2, N =0$. In this case,  one has the double-pole soliton solutions that can be obtained  using certain limiting process of \eqref{expression_u}, \cf \cite{GShosoliton}. 
  In general, the double-pole soliton solutions on the full-line   associated with the scattering data   $  \{\lambda_1;c_1, c_2 \}$  can be  expressed as  
   \begin{equation}
     q(x,t)=2 \Delta_1 \left(\frac{i\Delta_1(\gamma^*_2\gamma_1^2+\gamma_2) +2(1+ |\gamma_1|^2)\gamma_1}{\Delta_1^2|\gamma_2|^2+(1+|\gamma_1|^2)^2}\right)\,,
   \end{equation}
   where $\Delta_1=-2\,\text{Im}(\lambda_1)$, 
$
    \gamma_1: =c_1 e^{-2i\lambda_1(x+2\lambda_1t)}$, $ \gamma_2:= (c_2-2i(x+4t \lambda_1))\gamma_1 $,   and  $c_1, c_2$ as the norming constants associated to $\lambda_1$ can be freely chosen with $c_1\neq 0$. For the half-line problem, let $\lambda_1 = \beta_1^*$. Then,  one obtains a double-pole soliton completely absorbed by the boundary as displayed in Figure \ref{plot_2soliton_2poles}.   
   \begin{figure}[htp]
    \centering
    \includegraphics[width=6cm]{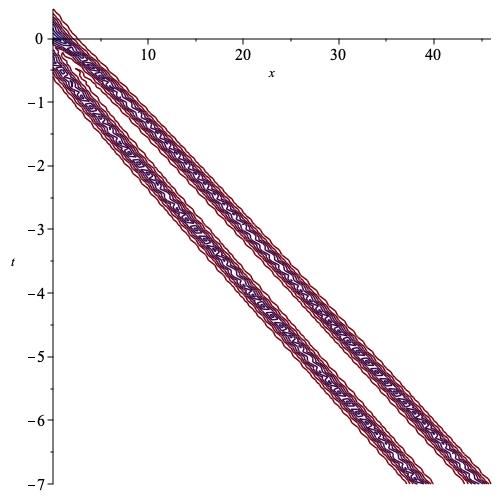}
    \hspace{2cm}
    \includegraphics[width=7cm]{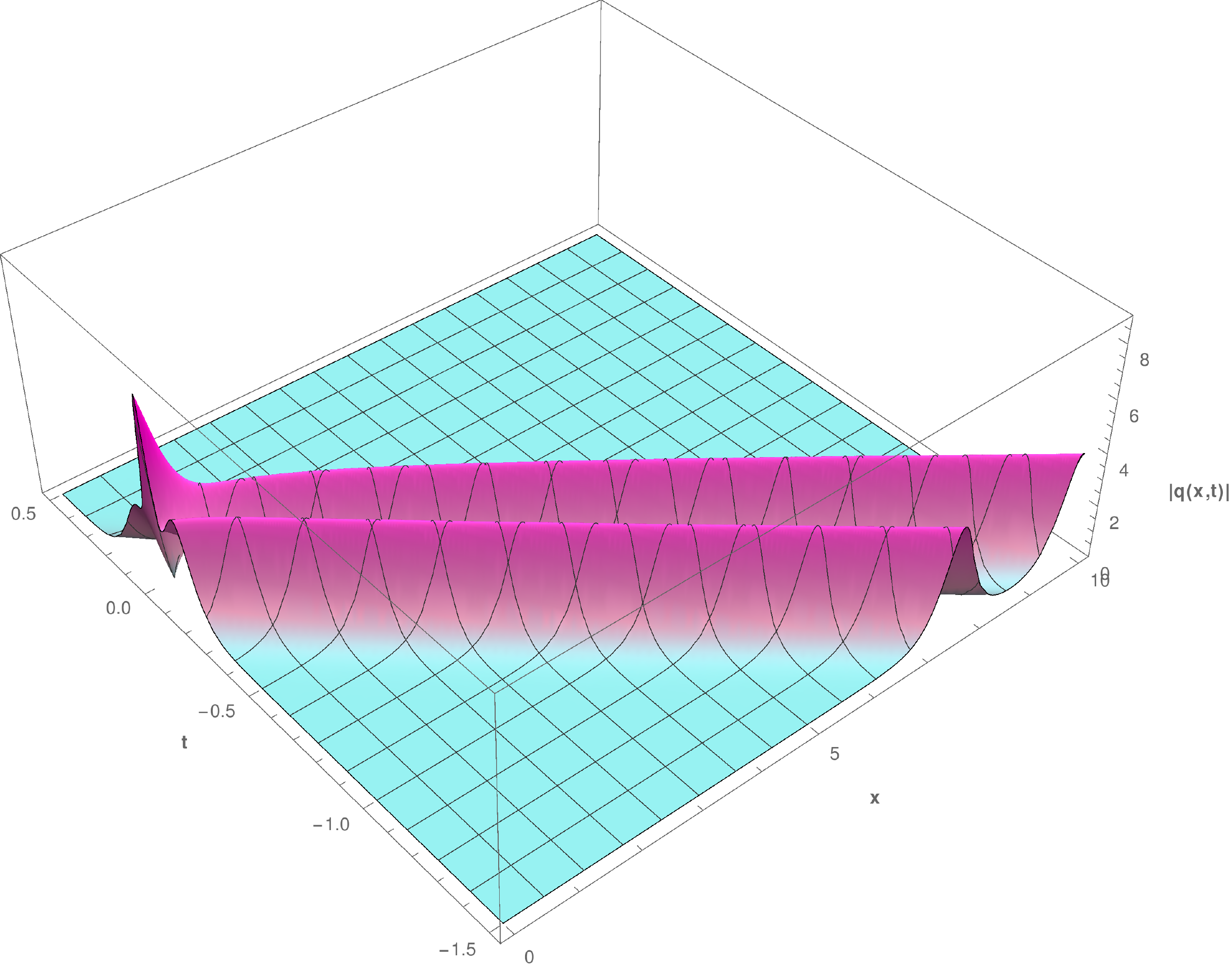}
    \caption{2D contour plots (on the left) and 3D plot (on the right)  of $|q(x,t)|$ corresponding to a double-pole soliton with time-dependent boundary conditions of $\cN=2$. One has  $\beta_1=3/2-2i$. The scattering data are $\{\beta_1^*; c_1= 2, c_2=0\}$. }    \label{plot_2soliton_2poles}
\end{figure}

Similarly, for  $\varphi(\lambda) =(\frac{\lambda-\beta_{1}}{\lambda-\beta^*_{1}})^2$, setting $ \lambda_1 = -\beta_1$ would provides a double-pole soliton completely emitted by the boundary. Other choices of $\varphi(\lambda)$ can be  treated  similarly as in Case I.

\end{description}

  

\section{Hamiltonian formalism\label{sec:Ham}}\label{Ham}

In this section, we present a formula for the conserved charges of the model with our time-dependent boundary conditions. The derivation is skipped as it is similar to the one given in \cite{CCD}. We then turn to Hamiltonian aspects of our model, restricting our attention to the continuous scattering data only and to real spectral parameters. This allows us to derive a classical version of the boundary algebra \cite{bound-alg} that was used originally in the context of the quantum nonlinear Schr\"odinger equation. 

\subsection{Integrability and conserved charges \label{sec:integrability}}

We compute the conserved quantities of the NLS model with boundary conditions dictated by the 
$K$-matrix given in \eqref{eq:Kex}. We introduce the function
\beq
G(x,t;\lambda) =\sum_{n=1}^\infty\frac{G^{(n)}(x,t)}{(2i\lambda)^n}
\eeq
whose coefficients are defined recursively as
\beq
G^{(1)}=q^*\,,\qquad G^{(n+1)}=(G^{(n)})_x+q\sum_{k=1}^{n-1}G^{(k)}G^{(n-k)} \,.
\eeq
By a standard argument, the quantities $q G^{(2n-1)}$ are the densities for the conserved charges of the model on the full-line. Here, it is natural to consider their half-line analogs
\begin{eqnarray}
I^{(2n-1)}&=&\int_{0}^\infty (q G^{(2n-1)})\,dx \,.
\end{eqnarray}
For instance, $I^{(1)}$ counts the number of solitons on the half-line (in appropriate units) and $I^{(3)}$ counts the energy.
However, the quantities $I^{(2n-1)}$ are not conserved in time. Instead, one can show as in \cite{CCD} that the following holds
\begin{eqnarray}
	\partial_tI^{(2n-1)}(t)=\partial_t\cK^{(2n-1)}(t)\,,
\end{eqnarray}
where $\cK^{(n)}(t)$ is given by
\begin{eqnarray}
\label{def_bd_charges}
\sum_{n=1}^\infty\frac{{\cal K}^{(n)}(t)}{(2i\lambda)^n}
    &=&\frac{1}{2}\ln\big(K_{11}(t;\lambda)+K_{12}(t;\lambda)G(0,t,\lambda)\big)\,.
\end{eqnarray}
Thus, the conserved quantities of the model with time-dependent boundary conditions are given by
\begin{eqnarray}
	\mathcal{I}^{(2n-1)}=I^{(2n-1)}(t)-\cK^{(2n-1)}(t)\,,
\end{eqnarray}
In \eqref{def_bd_charges}, $K_{ij}(t;\lambda)$ are the entries of the matrix $K(t;\lambda)$.
For the $K$-matrix given in \eqref{eq:Kex}, the first two non trivial conserved quantities reads (up to additive constant terms)  
\begin{equation}
 \begin{aligned}
& \cI^{(1)}=\int_0^{\infty}|q(x,t)|^2 dx  + f_{2\cN-1},\\
& \cI^{(3)}=\int_0^{\infty}\big( |q(x,t)|^4-|q_x(x,t)|^2\big)  dx -4f_{2\cN-3}+ \frac{2f_{2\cN-1}}{3}( 3|q(0,t)|^2+2f_{2\cN-1}^2+6f_{2\cN-2}) \,
\end{aligned}
\end{equation}
where $f_{-1}=0$, by convention. Recall that the functions $f_j$ depend on $t$ in general.

The conserved quantity $\cI^{(3)}$ is linked to the NLS Hamiltonian with a boundary term associated to the chosen $K$-matrix (see Section \ref{sec:PB}).
In the following, we provide the explicit form of this Hamiltonian (and also $\cI^{(1)}$) for the $K$-matrix studied in Section \ref{sec:Km} in the cases $\cN=1$ and $\cN=2$.

\paragraph{Example $\mathbf{\cN=1}$.} The first two conserved quantities read 
\begin{align}
& \cI^{(1)}=\int_0^{\infty}|q(x,t)|^2 dx  + f_1,\\
& \cI^{(3)}=\int_0^{\infty}\big(|q(x,t)|^4- |q_x(x,t)|^2\big)  dx  + \frac{2f_1}{3}\big(|q(0,t)|^2-2(\beta_1+\beta_1^*)^2 +2|\beta_1|^2\big)\,,
\end{align}
up to additive constant terms $ i(\beta_1-\beta_1^*) $ and  $-\frac{4i}{3}\big(\beta_1^3-(\beta_1^*)^3\big)$, respectively.
 The Hamiltonian of this model is chosen as $H=-\cI^{(3)}$ (as already given in \cite{ZAMBON}), such that the bulk equation of motion \eqref{eq:NLS} and the boundary condition \eqref{eq:bc1} are recovered, see Section \ref{sec:PB} below. 

\paragraph{Example $\mathbf{\cN=2}$.} In this case, the first two conserved quantities become
\begin{align}
& \cI^{(1)}=\int_0^{\infty}|q(x,t)|^2 dx  + f_3\,,\\
& \cI^{(3)}=\int_0^{\infty}\big( |q(x,t)|^4-|q_x(x,t)|^2\big)  dx -4f_1+ \frac{2f_3}{3}( 3|q(0,t)|^2+2f_3^2+6f_2) \,,
\end{align}
up to additive constant terms $i(\beta_1-\beta_1^*+\beta_2-\beta_2^*)$ and $-\frac{4i}{3}\big(\beta_1^3-(\beta_1^*)^3+\beta_2^3-(\beta_2^*)^3\big)$.

\subsection{Poisson structure\label{sec:PB}}

In this section, as required in the Hamiltonian formalism, time dependence arises as a flow on phase space and quantities in Poisson brackets do not depend on it explicitly. In other words,  The Poisson brackets are computed at equal time. Thus, we drop the variable $t$ in the quantities for which we compute the Poisson brackets. We
use the auxiliary space notation which is standard when dealing with classical $r$-matrix calculations. The reader is referred to \cite{FT} for details.

The bulk Poisson bracket between the fields 
\begin{eqnarray}
\{q(x)\,,\, q^*(y)\} &=& i\,\delta(x-y)\,\qquad \text{for $x,y>0$,}
\end{eqnarray}
can be recovered from the following Poisson bracket of the Lax matrix $U(x;\lambda)$:
\begin{eqnarray}
\{U(x;\lambda)\underset{'}{\otimes}U(y;\mu)\} = \delta(x-y)\,[U_1(x;\lambda)+U_2(y;\mu), r_{12}(\lambda-\mu)]\,,
\end{eqnarray}
where   $U_1 =  U\otimes 1\!\!1 $ and $U_2 = 1\!\!1\otimes U$, and we have  the usual classical $r$-matrix of the NLS model in the form (the subscripts $1$ and  $ 2$ refer to the auxiliary space notation)
\begin{eqnarray}
r_{12}(\lambda)=\frac1{2\lambda}\,P_{12}\,,\qquad \text{with }\
P_{12}=\begin{pmatrix} 1 & 0 &0 &0 \\ 0& 0& 1& 0\\ 0& 1& 0& 0\\ 0& 0& 0& 1\end{pmatrix}. 
\end{eqnarray}


Following the general approach on classical integrable systems with boundary, we obtain the Poisson structure at $x=0$
from the classical reflection equation \cite{sklyanin1988boundary,ACC} 
\begin{equation}
\begin{aligned}
\{K(\lambda)\underset{'}{\otimes}K(\mu)\} &= r_{12}(\lambda-\mu)\,K_1(\lambda)\,K_2(\mu) - K_1(\lambda)\,K_2(\mu)\,r_{12}(\lambda-\mu)
\\
&\quad -K_1(\lambda)\,r_{12}(\lambda+\mu)\,K_2(\mu) +K_2(\mu)\,r_{12}(\lambda+\mu)\,K_1(\lambda)\,,
\end{aligned}
\end{equation}
where $K_1 = K\otimes 1\!\!1 $ and $K_2 = 1\!\!1\otimes K$.  Writing 
\begin{eqnarray}
K(\lambda) &=& \chi_0(\lambda)\,1\!\!1 +\chi_3(\lambda)\,\sigma_3 +\chi_+(\lambda)\,\sigma_+ +\chi_-(\lambda)\,\sigma_-\,,
\end{eqnarray}
where
\[\sigma_+ =\bma 0 & 1 \\ 0 &0 \ema, \quad \sigma_- =\bma 0 & 0 \\ 1 &0 \ema \,,\]
we get as non-vanishing Poisson brackets
\begin{subequations}
\begin{eqnarray}
\{\chi_0(\lambda),\chi_3(\mu)\} &=& \frac{\lambda}{2(\lambda^2-\mu^2)}\Big( \chi_+(\lambda)\,\chi_-(\mu)
-\chi_-(\lambda)\,\chi_+(\mu)\,\Big)
\\
\{\chi_0(\lambda),\chi_\pm(\mu)\}  &=& \frac{\pm\lambda}{\lambda^2-\mu^2}\Big( 
\chi_3(\lambda)\,\chi_\pm(\mu)-\chi_\pm(\lambda)\,\chi_3(\mu)\,\Big)
\\
\{\chi_3(\lambda),\chi_\pm(\mu)\}  &=& \frac{\pm1}{\lambda^2-\mu^2}\Big( \lambda\,\chi_0(\lambda)\,\chi_\pm(\mu)
-\mu\,\chi_0(\mu)\,\chi_\pm(\lambda)\,\Big)
\\
\{\chi_+(\lambda),\chi_-(\mu)\}  &=& \frac{2}{\lambda^2-\mu^2}\Big( 
\lambda\,\chi_3(\mu)\,\chi_0(\lambda)-\mu\,\chi_0(\mu)\,\chi_3(\lambda)\,\Big)
\end{eqnarray}
\end{subequations}
Using the expansion \eqref{eq:Kex}, it allows to determine the Poisson brackets of the functions $f_k$, $g_\ell$ and $q(0)$, as we shall see in an example.

\paragraph{Example $\mathbf{\cN=1}$.} In that case, $\chi_0(\lambda)=\lambda^2+f_0$, $\chi_3(\lambda)=i\lambda f_1$,
$\chi_+(\lambda)=i\lambda q(0)$ and $\chi_-(\lambda)=i\lambda q^*(0)$, where the (irrelevant here) normalisation of $K$ has been omitted. It leads to the non-vanishing Poisson brackets
\begin{equation}\label{PB:N=1}
\{q(0),q^*(0)\} = -2 i f_1 \mb{;} \{f_1,q(0)\} = - i q(0) \mb{and} \{f_1,q^*(0)\} =  i q^*(0)\,.
\end{equation}
The  last two Poisson brackets can be consistently deduced from the first one and the definition of $f_1=-|q(0)|^2-(\beta_1-\beta_1^*)^2$.
Using the Poisson bracket structure \eqref{PB:N=1}
and writing the Hamiltonian as
\begin{equation}
H=-\int_0^{\infty}\big(q_{xx}(x,t)q^*(x,t)+|q(x,t)|^4\big)  dx -q_x(0,t)q^*(0,t)
 - \frac{2f_1}{3}\big(|q(0,t)|^2-2(\beta_1+\beta_1^*)^2 +2|\beta_1|^2\big)\,
\end{equation}
we recover the equation of motion \eqref{eq:NLS} and the boundary condition 
\eqref{eq:bound-N=1} through the relation $q_t=\{H,q\}$ for $x>0$ and $x=0$ respectively.


\subsection{Poisson structure of the scattering data\label{sec:Bound_alg}}

We now turn to the derivation of the Poisson structure of the continuous scattering data in a form suitable for comparison with a quantum algebra, called boundary algebra, which was introduced in \cite{bound-alg} to describe integrable quantum field theories with a boundary. We define
\beq
\KK(\lambda) = \Gamma(\lambda) \,K_+(\lambda)^{-1}\,,
\eeq
where $\Gamma(\lambda)$ is defined in \eqref{GGGG} with its entries obeying the symmetry relations \eqref{sym_A}.
When $\lambda$ is real, the functions $g_A$ and $g_B$ (see \eqref{eq:GGG}) obey
\beq\label{prop:gAgB}
g_A(\lambda)\,g_A(-\lambda)=1=g_B(\lambda)\,g_B(-\lambda)\,,
\quad g_A(\lambda)\,g_A^*(\lambda)=1=g_B(\lambda)\,g_B^*(\lambda).
\eeq
From \eqref{eq:G-matrix}, we deduce
\beq\label{eq:Skly}
\KK(\lambda) = \frac1{\vph(\lambda)}\,\begin{pmatrix} {A}(\lambda) & -u(\lambda)\,{B}^*(\lambda^*) \\ {B}(\lambda) & u(\lambda)\,{A}^*(\lambda^*) \end{pmatrix}
\mb{with}  u(\lambda)=-g_A(\lambda)\,g_B(\lambda).
\eeq
Following Sklyanin \cite{sklyanin1988boundary}, the quantity $\KK(\lambda)$ satisfies the following classical limit of the (quantum) reflection algebra
\beq
\begin{aligned}
\{ \KK(\lambda_1) \underset{'}{\otimes} \KK(\lambda_2)\} &=  \KK_2(\lambda_2)\,\KK_1(\lambda_1)\,r_{21}(-\lambda_2,-\lambda_1) 
-r_{12}(\lambda_1,\lambda_2)\,\KK_1(\lambda_1)\,\KK_2(\lambda_2) \\
&\quad+\KK_2(\lambda_2)\,r_{12}(\lambda_1,-\lambda_2)\,\KK_1(\lambda_1)
-\KK_1(\lambda_1)\,r_{21}(\lambda_2,-\lambda_1)\,\KK_2(\lambda_2)\,,
\end{aligned}
\eeq
where $r_{21}(\lambda_2,\lambda_1)=P_{12}r_{12}(\lambda_2,\lambda_1)P_{12}$. This leads to the following Poisson brackets:
\beq\label{PB-AB}
\begin{aligned}
\{{A}(\lambda_1)\,,\,{A}(\lambda_2)\} &= 0\quad;\quad
\{{A}(\lambda_1)\,,\,{A}^*(\lambda_2)\} = 0\quad;\quad
\{{B}(\lambda_1)\,,\,{B}(\lambda_2)\} = 0\\
\{{A}(\lambda_1)\,,\,{B}(\lambda_2)\} &= \left(\frac{-{1/2}}{\lambda_1-\lambda_2} - \frac{{1/2}}{\lambda_1+\lambda_2}\right) {A}(\lambda_1)\,{B}(\lambda_2)
\\
&\qquad+\frac{i\pi}{2}\Big(\delta(\lambda_1-\lambda_2) +\delta(\lambda_1+\lambda_2)\Big){A}(\lambda_1)\,{B}(\lambda_2) \\
\{{A}(\lambda_1)\,,\,{B}^*(\lambda_2)\} &= {\frac12}\left(\frac{1}{\lambda_1-\lambda_2} - \frac{1}{\lambda_1+\lambda_2}\right) 
\,{A}(\lambda_1)\,{B}^*(\lambda_2)\\
&\qquad-\frac{i\pi}{2}\Big(\delta(\lambda_1-\lambda_2) 
+\delta(\lambda_1+\lambda_2)\Big){A}(\lambda_1)\,{B}^*(\lambda_2)\\
\{{B}(\lambda_1)\,,\,{B}^*(\lambda_2)\} &= -i\pi\,\delta(\lambda_1-\lambda_2) \,{A}(\lambda_1)\,{A}^*(\lambda_1)
+\frac{i\pi}{u(\lambda_2)}\,\delta(\lambda_1+\lambda_2)\,{A}(\lambda_1)\,{A}(\lambda_2)
\end{aligned}
\eeq

We introduce the following functions 
\beq
\begin{aligned}
&{C}(\lambda)=\ff\left(\frac{|B(\lambda)|^2}{|A(\lambda)|^2}\right)\,\frac{B(\lambda)}{A(\lambda)} \mb{;} 
{C}^*(\lambda)= \ff\left(\frac{|B(\lambda)|^2}{|A(\lambda)|^2}\right)\,\frac{B^*(\lambda)}{A^*(\lambda)} \\
&R(\lambda) = \frac{g_B(\lambda)}{g_A(\lambda)}\,\frac{A(\lambda)}{A^*(\lambda)}\mb{;} 
{N}(\lambda)= {C}^*(\lambda){C}(\lambda)
\end{aligned}
\eeq
where $\ff(x)$ is a real function obeying the differential equation
\beq \label{eqdiff:f}
\big(1+\,x\big)\,\Big(2x\,\ff'(x)+\ff(x)\Big)\ff(x)=2.
\eeq
A solution for $\ff$ that is well-defined for $x\ge 0$ is given by $\ff(x)=\sqrt{\frac{\ln(1+ x)}{ x}}$. In ${C}(\lambda)$ and ${C}^*(\lambda)$, we have the combinations $\frac{B(\lambda)}{A(\lambda)}$ and $\frac{B^*(\lambda)}{A^*(\lambda)}$ similar to the familiar reflection coefficients appearing in ISM on the line. The extra normalisation factor $\ff\left(\frac{|B(\lambda)|^2}{|A(\lambda)|^2}\right)$ is introduced here so that ${C}(\lambda)$ and ${C}^*(\lambda)$ have Poisson brackets that can directly be interpreted as the classical limit of the corresponding relation in the boundary algebra. In the latter context, the corresponding (quantum) operators ${C}(\lambda)$ and ${C}^*(\lambda)$ have the interpretation of creation and annihilation operators for asymptotic states with momentum $\lda$. The quantum counterpart of $R(\lda)$ is the boundary operator whose action on the vacuum state of the theory yields the reflection matrix and labels the (Fock) representation of the boundary algebra. Finally the quantum counterpart of $N(\lda)$ is the number operator which counts the number of particles created by ${C}^*(\lambda)$.

Due to the symmetry relations \eqref{sym_A}, we have
\beq
 C(-\lambda) =\frac{g_B(\lambda)}{g_A(\lambda)}\, \frac{A(\lambda)}{A^*(\lambda)}\,  C(\lambda)\equiv R(\lambda)\, C(\lambda)\,,
\eeq
in agreement with the interpretation of $R(\lda)$ as a classical boundary ``operator''.
Remark that due to the relations \eqref{prop:gAgB}, we have the property 
\beq
R(\lambda)\,R^*(\lambda)=1\mb{and}R(\lambda)\,R(-\lambda)=1\,.
\eeq

Then, in this new basis, we get 
\beq
\begin{aligned}
\{{C}(\lambda_1)\,,\,{C}(\lambda_2)\} &= \frac{1}{\lambda_1-\lambda_2}\,{C}(\lambda_1)\,{C}(\lambda_2)\\
\{{C}(\lambda_1)\,,\,{C}^*(\lambda_2)\} &= \frac{-1}{\lambda_1-\lambda_2}\,{C}(\lambda_1)\,{C}^*(\lambda_2)
+2i\pi\Big(\delta(\lambda_1-\lambda_2)+\delta(\lambda_1+\lambda_2)\,R^*(\lambda_1)\Big)\ \\
\{{C}(\lambda_1)\,,\,R(\lambda_2)\} &= \left(\frac{-1}{\lambda_1-\lambda_2}+\frac{1}{\lambda_1+\lambda_2}\right)\,{C}(\lambda_1)\,R(\lambda_2) \\
\{{C}^*(\lambda_1)\,,\,R(\lambda_2)\} &= \left(\frac{1}{\lambda_1-\lambda_2}+\frac{1}{\lambda_1+\lambda_2}\right)\,{C}^*(\lambda_1)\,R(\lambda_2)\\
\{R(\lambda_1)\,,\,R(\lambda_2)\} &= 0 \\
\{{N}(\lambda_1)\,,\,{C}(\lambda_2)\} &= -2i\pi\,{C}(\lambda_2)\Big(\delta(\lambda_1-\lambda_2) 
+\delta(\lambda_1+\lambda_2)\Big)\\
\{{N}(\lambda_1)\,,\,{C}^*(\lambda_2)\} &=2 i\pi\,{C}^*(\lambda_2)\Big(\delta(\lambda_1-\lambda_2) 
+\delta(\lambda_1+\lambda_2)\Big)\\
\{{N}(\lambda_1)\,,\,{N}(\lambda_2)\} &= 0\quad;\quad \{{N}(\lambda_1)\,,\,R(\lambda_2)\}=0\,.
\end{aligned}\eeq

What we have achieved is the derivation, for the first time, of the classical version (in the sense of the classical limit explained in \cite{sklyanin1988boundary}) of the boundary algebra introduced in \cite{bound-alg}, with the choice,
\beq\label{eq:matS-NLS}
s_{12}(\lambda_1,\lambda_2) = \frac{-1}{\lambda_1-\lambda_2}\,.
\eeq
Let us emphasize the choice of terminology whereby (classical) boundary algebra, following \cite{bound-alg}, should not be confused with (classical) reflection algebra, as presented in \cite{sklyanin1988boundary}. The former allows to construct solutions to integrable field theories, while the solutions to the latter provide reflection matrices describing integrable boundary conditions within these field theories. 

\section{Conclusions}\label{Ccl}

We presented an in-depth study of the effect of absorption and/or emission of solitons by a boundary described by a time-dependent reflection matrix. The results and formulas imply that a rich variety of phenomena can occur which include reflection of (higher order) solitons together with absorption/emission controlled by the boundary parameters. We illustrated the simplest key scenarios. These findings show that the original results found in \cite{CCD} are part of a large family of boundary phenomena. The absorption/emission phenomenon of solitons is a signature of time-dependent integrable boundary conditions and is absent in the well-known case of Robin boundary conditions, as shown in \cite{CCD}. In the present paper, we opted to take advantage of the results of \cite{Zist-half-line} to construct solutions instead of using the nonlinear mirror image as was done in \cite{CCD}. Although this is not reported explicitly here, we did compare the two methods and check their consistency. This involved the construction of the appropriate Darboux-B\"acklund matrix $L(x,t;\lda)$ and showing that it correctly interpolates between $K_+$ as $x\to \infty$ and $K(t;\lambda)$ at $x=0$. 

Our results raise the question of classifying possible time-dependent reflection matrices for other integrable models and investigate if similar absorption/emission effects take place. This would require solving the analog of \eqref{eq:c1} for the Lax matrix $V$ associated to the desired model. We expect the phenomenon to be rather generic but with the details being model dependent. Perhaps the simplest next model to investigate would be the vector nonlinear Schr\"odinger equation. Indeed, the possible phenonema could be even richer due to the extra degrees of freedom in this model which allow for more intricate time-dependent boundary conditions, as discussed in the recent work \cite{Xia2}. Hamiltonian and Lax pair aspects of vector NLS with time-independent integrable boundary conditions were presented earlier in \cite{DFR,Doi}.

The results of Section 4 could provide the basis for quantization of our construction following for instance the ideas of \cite{GLM1,GLM2,MRS}. So far only time-independent (Robin) boundary conditions have been quantized along these lines, for  free fields or for the NLS equation on the half-line. The quantization of our time-dependent boundary conditions appears exciting but rather challenging, due to their high nonlinearity. The phenomenon of absorption/emission displayed here is particularly puzzling from the quantization point of view.  It might fall instead into the remit of the more general Reflection-Transmission algebras introduced in \cite{MRS2} and used in \cite{CMR1,CMR2} to study the quantum NLS with an impurity. This is unclear at this stage and deserves further investigation.

\section*{Acknowledgments}
N. Cramp\'e is supported by the international research project AAPT of the CNRS. C. Zhang is supported by NSFC (No. 11875040, 12171306). 
\appendix 

\section{Proof of Prop. \ref{prop:main}}
\label{app:prf1}The proof takes advantage of the ISM for the NLS model on the half-line from the double-row monodromy matrix \cite{Zist-half-line}. In this setting, the direct scattering process transforms the initial-boundary conditions into the scattering functions $A(\lambda)$, $B(\lambda)$ that are the entries of the double-row monodromy matrix $\Gamma(t;\lambda)$, \cf \eqref{eq:G-matrix}. The inverse part of the ISM can be formulated as a Riemann-Hilbert problem which can be shown to be equivalent to the half-line NLS equation subject to the boundary conditions determined by Prop. \ref{prop:Kexpansion}. Details can be found in  \cite{Zist-half-line}. Here, we simply formulate the results in the direct scattering part. This relies on the following scattering system
\begin{equation}\label{eq:sc1}
  Y (x,t;\lambda) =  X (x,t;\lambda)\Gamma(t;\lambda)\,,\quad \lambda \in \RR\,,
\end{equation}
where  $\Gamma(t;\lambda)$ takes the form \eqref{eq:G-matrix}, and $X, Y$ are $2\times 2$ matrix-valued functions playing the roles of the normalized {\it Jost solutions}. Let
\begin{equation}
  Y= \bma Y_{11}  & Y_{12} \\ Y_{21} & Y_{22} \ema \,,\quad   X= \bma X_{11}  & X_{12} \\ X_{21} & X_{22} \ema \,,  
\end{equation}
one can show that the first column of $Y$ and second column of $X$  are analytic and bounded in the upper half complex plane, while the second column of $Y$ and the first column of $X$ are analytic and bounded in the lower half complex plane  \cite{Zist-half-line}.
For simplicity, let $x=t=0$ so that the space and time dependences are omitted. This will not affect the relations among the discrete data.

It follows from \eqref{eq:sc1} that
\begin{equation}
  A(\lambda)  =\det \bma Y_{11}  &   X_{12}\\ Y_{21} & X_{22} \ema\,,\quad   A^*(\lambda^*)  =\det \bma X_{11}  & Y_{12} \\ X_{21} & Y_{22} \ema\,. 
\end{equation}
If
  \begin{equation}\label{eq:norm1}
\bma  X_{12} \\ X_{22}\ema\Big\vert_{\lambda= \mu_\ell} = c_\ell    \bma  Y_{11} \\ Y_{21}\ema\Big\vert_{\lambda= \mu_\ell} \,,
  \end{equation} with $\mu_\ell$ being a complex number with positive imaginary part, then $  A(\mu_\ell) = 0$, or $\mu_\ell$ is a zero of $A(\lambda)$. This means there is a {\it bound state} associated to the discrete spectrum $\mu_\ell$ for  the Lax system \eqref{eq:laxeqs}, and  $c_\ell$ is the associated norming constant. 
  A collection of $\{\mu_\ell; c_\ell\}\vert_{\ell=1,\dots, N}$, with distinct $\mu_\ell$ as simple zeros of $A(\lambda)$, will generate $N$-soliton solutions as in the case of full-line problem. 

  Now taking the second involution relation \eqref{eq:inv222} into account. First, assume $\mu_\ell$ does not coincide with any boundary parameter $\beta_j^*$ or $-\beta_j$, what we are showing is that there are simultaneously paired scattering data $\{  \widetilde{\mu}_\ell; \widetilde{c}_\ell \}\vert_{\ell = 1, \dots, N}$ obeying \eqref{sym_norming}. Due to the first formula in \eqref{sym_A}, if $A(\mu_\ell) =0$, then   $A(-\mu_\ell^*) =0$, and {\it vise versa}. This leads to a bound state  as
      \begin{equation}\label{eq:norm2}
  \bma  X_{12} \\ X_{22}\ema\Big\vert_{\lambda= -\mu^*_\ell} = \widetilde{c}_\ell  \bma  Y_{11} \\ Y_{21}\ema\Big\vert_{\lambda= -\mu^*_\ell}\,,  
  \end{equation}
with $\widetilde{c}_\ell$ to be determined. Taking $K_+(\lambda)$ in the form \eqref{K_split}, and using  \eqref{eq:inv222}, one has 
  \begin{equation}
    \bma  Y_{11} \\ Y_{21}\ema(\lambda) = \varphi(\lambda)K_+(-\lambda)\bma  X_{11} \\ X_{21}\ema(-\lambda)\,,\quad    \bma  Y_{12} \\ Y_{22}\ema(\lambda) = \frac{\kappa(\lambda)}{\varphi(\lambda)}K_+(-\lambda)\bma  X_{12} \\ X_{22}\ema(-\lambda) \,. 
  \end{equation}
This leads to 
  \begin{equation}
    \bma  Y_{11} \\ Y_{21}\ema\Big\vert_{\lambda= \mu_\ell} =\varphi(\mu_\ell) K_+(-\mu_\ell)\bma  X_{11} \\ X_{21}\ema\Big\vert_{\lambda=- \mu_\ell} =\varphi(\mu_\ell) K_+(-\mu_\ell)\bma  X^*_{22} \\ -X^*_{12}\ema\Big\vert_{\lambda=- \mu^*_\ell}\,,
  \end{equation}
  where the last equality follows from the involution relation \eqref{eq:invg}. Similarly, one has
  \begin{equation}
    \bma  X_{12} \\ X_{22}\ema\Big\vert_{\lambda= \mu_\ell} =    \frac{\varphi(-\mu_\ell)}{\kappa(-\mu_\ell)} K_+(-\mu_\ell)\bma  Y_{12} \\ Y_{22}\ema\Big\vert_{\lambda= -\mu_\ell} =\frac{\varphi(-\mu_\ell)}{\kappa(-\mu_\ell)} K_+(-\mu_\ell)\bma  -Y_{21}^* \\ Y_{11}^*\ema\Big\vert_{\lambda= -\mu^*_\ell}\,. 
  \end{equation}
Taking \eqref{eq:norm1} and \eqref{eq:norm2}  into account, one has the desired relation between $c_l$ and $\widetilde{c}_l$. This pairing mechanism  generates $2N$-soliton solutions on the full-line with opposite velocities, which leads to $N$ reflection solitons on the half-line by restricting $x\geq 0$. The boundary conditions can be understood as interactions among the $2N$ solitons. 

  If one or several zeros of $A(\lambda)$ coincide with the boundary parameters as shown in \eqref{expression_A}, then the associated norming constants can be freely chosen. This corresponds to the case where absorption/emission of solitons by the boundary take place.   

\section{Dressing of $K(t;\lambda)$}\label{Appendix_dressing}
In general, from Sklyanin's work \cite{sklyanin1988boundary}, one expects that a time-dependent (dynamical) reflection matrix such as the one we are working with should be of the factorised form
 \begin{equation}\label{factorised_K}
 	K(t,\lda)=T(t,-\lda)^{-1}\,N(\lda)\,T(t,\lda)
 \end{equation}
where $N(\lda)$ is a purely numerical, time-independent, reflection matrix and $T(t,\lda)$ can be viewed as a transition matrix and contains the time-dependence. This structure is clear in the works \cite{ZAMBON,CCD} which correspond to the case $\cN=1$, $\cM=0$ here. However, for convenience in the present work, we choose to work with $K(t,\lda)$ as in \eqref{eq:Kex} so it is not clear whether \eqref{factorised_K} holds in our case. The purpose of this Appendix is to show that this is the case by using the same idea as in \cite{CCD}, \ie by dressing repeatedly an appropriate matrix $N(\lda)$.
Two remarks are in order for such a purpose. First, the $K$-matrix, once assumed to be of the form \eqref{eq:Kex}, is uniquely defined by the relation \eqref{eq:c1}. 
Second, the factorised B\"acklund matrix can be constructed by dressing iteratively as follows
\begin{equation}
\cB(x,t;\lambda)\ \to\ \cL(-x,t;-\lambda)^{-1}\,\cB(x,t;\lambda)\,\cL(x,t;\lambda)\label{eq:back}\,.
\end{equation}
In \eqref{eq:back}, the Lax operator $\cL(x,t;\lambda)$ is defined as (up to a normalisation function):
\begin{equation}
	\label{form_L}
 \cL(x,t;\lambda) =  (\lambda+\rho)\,\sigma_3+ \begin{pmatrix} i\, h_0(x,t)& h_+(x,t) \\ h_+^*(x,t) & i\, h_0(x,t) \end{pmatrix}\,,
\end{equation}
where $h_0(x,t)$ is a real-valued function and $\text{det}\cL(x,t;\lambda)$ is independent of $x,t$, see eq. (4.11) in \cite{CCD}.  The latter fact, together with the explicit expression $\text{det}\cL(x,t;\lambda)=-(\lambda+\rho)^2- h_0(x,t)^2 -|h_+(x,t)|^2$, show that $\text{det}\cL(x,t;\lambda)$ is a polynomial in $\lambda$ of degree $2$ with real coefficients, which implies that the determinant can be factorised as $- (\lambda+\beta)(\lambda+\beta^*)$ for some $\beta\in\CC$. Thus the B\"acklund matrix is characterised by two real numbers which we call its {\it parameters}.

The first step of the recursion is given by the results of \cite{CCD}, see eq. (5.11) there. We have (in the $\cN=1$ case)
$K(t,\lda)=\cB_1(0,t;\lda)$ where $\cB_1(x,t;\lda)=L_1(-x,t,-\lda)^{-1}\,L_1(x,t;\lda)$ and $L_1(x,t;\lda)$ is exactly of the form \eqref{form_L}. This establishes 
\eqref{factorised_K} with $N(\lda)=\1$ and $T(t,\lda)=L_1(0,t;\lda)$. Noting that $K(t,\lda)=\cB_1(0,t;\lda)$ has the form \eqref{eq:Kex} thus establishes the first step of the recursion.

\medskip

Next, suppose we have a matrix $\cB_{\cN}(x,t;\lda)$ which, at $x=0$, is of the form \eqref{eq:Kex}, with real-valued functions $f_j(t)$, complex-valued functions $g_j(t)$ and $f_{2\cN}(t)=1$, and also of the form \eqref{factorised_K}, with $N(\lda)=\1$ and some $T_{\cN}(t,\lda)$. Following the dressing idea that led to $\cB_{1}(x,t;\lda)$, we dress it as in \eqref{eq:back} using the following normalised version of \eqref{form_L}
	\begin{equation}\label{eq:L}
		L(x,t;\lambda) = \frac{-1}{\lambda+\beta_{\cN+1}}\,\Big[ (\lambda+\rho)\,\sigma_3+ \begin{pmatrix} i\, h_0(x,t)& h_+(x,t) \\ h_+^*(x,t) & i\, h_0(x,t) \end{pmatrix}\Big]
	\end{equation}
	with $\rho$ a constant real number, $ h_0(x,t)$ real-valued function, and $\beta_{\cN+1}$ a  complex parameter given by
	\begin{equation}\label{det_facto}
		(\lambda+\rho)^2+ h_0(x,t)^2 +|h_+(x,t)|^2=(\lambda+\beta_{\cN+1})(\lambda+\beta^*_{\cN+1})
		\,.
\end{equation}
Recall that $\text{det}\cL(x,t;\lambda)$ is a constant polynomial in $\lda$ so this indeed defines a constant $\beta_{\cN+1}$.
It is easy to see that $B_{\cN+1}(0,t;\lambda)=L(0,t;-\lambda)^{-1}\,B_{\cN}(0,t;\lambda)\,L(0,t;\lambda)$ has also the form \eqref{eq:Kex}, with $\cN\to\cN+1$. By construction, it also has the factorised form \eqref{factorised_K} with $T_{\cN+1}(t,\lda)=T_{\cN}(t,\lda)\,L(0,t;\lambda)$. Finally, by the properties of the dressing method, it satisfies \eqref{eq:c1} and therefore it is equal to $K(t,\lda)$ (by the uniqueness property recalled above). This shows the desired result at $\cN+1$. It also shows that the boundary parameters arise as the parameters that are known to be associated with each NLS B\"acklund matrix $L(x,t;\lda)$. Summarising our discussion, we have
\begin{proposition}
The time-dependent reflection matrix $K(t,\lda)$ in \eqref{eq:Kex} possesses the factorised form  	$$K(t,\lda)=T(t,-\lda)^{-1}\,N(\lda)\,T(t,\lda)$$
where $N(\lda)=\1$ and $T(t,\lda)$ is a product of successive NLS B\"acklund matrices evaluated at $x=0$,  with the parameters fixed by the boundary parameters by requiring that their determinant factorises as in \eqref{det_facto}.
\end{proposition}

\end{document}